\documentclass[usenatbib,useAMS]{mn2e}
\usepackage{times}
\usepackage{epsfig}

\title[Microlensing towards the LMC: self lensing for OGLE-II and OGLE-III]
{Microlensing towards the LMC: self lensing for OGLE-II and OGLE-III}

\author[S.~Calchi~Novati \& L.~Mancini]
{S.~Calchi Novati$^{1,2}$ 
\thanks{E-mail:novati@sa.infn.it} and
L.~Mancini$^{1,2,3}$\\
$^{1}$ Dipartimento di Fisica ``E. R. Caianiello'', Universit\`a di Salerno, 
Via Ponte don Melillo, 84084 Fisciano (SA), Italy\\
$^{2}$ Istituto Internazionale per gli Alti Studi Scientifici (IIASS), Vietri Sul Mare (SA), Italy\\
$^{3}$ Dipartimento di Ingegneria, Universit\`a del Sannio, Benevento, Italy
}
\begin{document}

\date{date}

\pagerange{\pageref{firstpage}--\pageref{lastpage}} \pubyear{2011}

\maketitle

\label{firstpage}

\begin{abstract}
We present an analysis of the results of the OGLE-III
microlensing campaign towards the Large Magellanic Cloud (LMC).
We evaluate for all the possible lens populations
along the line of sight the expected microlensing quantities,
number of events and duration. In particular we consider
lensing by massive compact halo objects (MACHOs) in the dark matter
haloes of both the Milky Way (MW) and the LMC,
and ``self lensing'' by stars in the LMC bar and disc,
in the MW disc and in the stellar haloes of both the LMC and the MW.
As a result we find that the self-lensing signal is able
to explain the 2 OGLE-III microlensing candidates.
In particular, we estimate the expected MW disc signal
to be almost as large as that from LMC stars and able, by itself,
to explain the observed rate.
We evaluate a 95\% CL \emph{upper} limit for $f$,
the halo mass fraction in form of MACHOs,
in the range 10-20\% for $(10^{-2}-0.5)~\mathrm{M}_\odot$,
and $f=24\%$ for $1~\mathrm{M}_\odot$ (below 10\% in this full range,
and in particular below 5\% for $(10^{-2}-0.1)~\mathrm{M}_\odot$) 
for the Bright (All) samples of source stars.
Furthermore, we find that these limits do not rise much
even if we assume the observed events \emph{are} MACHOs.
For the All sample we also evaluate a rather significant constraint
on $f$ for larger values of the MACHO mass, in particular $f\sim 50\%$ (95\% CL) for
$100~\mathrm{M}_\odot$, to date the stronger bound
coming from microlensing analyses in this mass range.
Finally, we discuss these results in the framework of
the previous observational campaigns towards 
the LMC, that of the MACHO and the EROS collaborations,
and we present a joint analysis of the OGLE-II and the OGLE-III campaigns.
\end{abstract}

\begin{keywords}
Gravitational lensing: micro; Galaxy: halo; Galaxy: structure; dark matter
\end{keywords}

\section{Introduction}

Microlensing has been originally proposed by \cite{pacz86}
as a tool to detect dark matter in form of (faint) massive compact halo
objects (MACHOs) in the galactic haloes. Because of the small
optical depth (we refer to \cite{mao08} and references therein
for an introduction to microlensing), dense stellar field has
to be monitored to increase the rate of events. The more
nearby available targets to look for MACHOs within the Galactic
halo are the Magellanic Clouds (the LMC and the SMC),
that have been therefore the objects of the first
microlensing campaigns by the MACHO \citep{macho93}, 
the EROS \citep{eros93} and the OGLE \citep{ogle93} collaborations
and for which several results have been reported so far
(for a recent review we refer to \citealt{moniez10}).
Meanwhile microlensing has moved on and currently
is an established tool of investigation over
a broad range of astrophysical issues, with
one of the main focuses being currently 
the search and the characterization of extra solar
planets \citep{smc2011}.

As for microlensing searches towards the Magellanic Clouds,
previous analyses  are in agreement to exclude MACHOs 
as a viable dark matter candidate for masses
below $\approx (10^{-1}-10^{-2})~\mathrm{M}_\odot$.
However, a relevant discrepancy still exists
as for a possible population of compact halo objects in the mass range
$(0.1-1)~\mathrm{M}_\odot$. The
MACHO collaboration claimed for a halo mass fraction
in form of MACHOs of about $f\sim 20\%$ out 
of observations of 13-17 microlensing candidate 
events towards the LMC \citep{macho00}, a result
more recently confirmed by \cite{bennett05},
who in particular confirmed the microlensing nature of
10-12 out of the original set of 13 candidate events.
On the other hand, the EROS \citep{eros07} and  
OGLE-II \citep{ogle09,ogle10_smc}, out of observations
towards both the LMC and SMC, concluded 
by putting extremely severe \emph{upper} limits
on the MACHO contribution also in this mass range
(in particular, the EROS collaboration reported
an upper limit $f=8\%$ for $0.4~\textrm{M}_\odot$ MACHOs).
Finally, we recall that results on MACHOs
through microlensing searches have been reported also
from observational campaigns towards M31 \citep{grg10}.

A source of debate for these results
is the intrinsic \emph{nature} of the reported microlensing events. 
Indeed, in order to draw meaningful insights into the
contribution of compact halo dark matter objects
it is essential first to estimate accurately all 
the possible contributions due to (luminous) lenses belonging to known 
populations located along the line of sight.
We refer to this possibility (first addressed by \cite{sahu94} 
and thereby the objects of several investigations,
among which \citealt{gyuk00,jetzer02})
as, broadly speaking, ``self lensing'',
to indicate any lens population that is not composed by MACHOs.
A possibly non exhaustive list includes lenses belonging
to the luminous components of the LMC
which act also as sources (the disc and the bar),
the disc of the MW and the somewhat more
elusive stellar haloes for both the MW and the LMC.

Hints on the nature of single specific
events require additional informations necessary
to break the intrisic degeneracies
within the lensing parameter space. 
These may come from observed 
features along the microlensing light curves,
as in binary systems, or by new independent measurements.
This issue has been addressed for a few Magellanic Clouds
events, and in particular for two SMC events both
found to be attributed more likely to self lensing:
the caustic crossing binary event  MACHO 98-SMC-1
(analysed for instance by \citealt{macho99,rhie99}),
and MACHO 97-SMC-1, a long duration event
for which a spectroscopic analysis of the source has been carried out by \cite{sahu98}.
On the other hand, the halo lensing solution is strongly favored for OGLE-2005-SMC-001
thanks to a space-based parallax measurement \citep{dong07}.
The SMC is any case somewhat peculiar with respect
to the LMC, both for its intrinsic density distribution
and orientation, and a global analysis for the events 
detected along this line of sight is still missing.
We also recall the LMC event MACHO-LMC-5,
whose lens has been directly observed by means 
of the HST \citep{alcock01b} and finally
aknowledged to be a MW disk M-dwarf lying a distance
at about $580~\mathrm{pc}$  \citep{gould04,gould04a,drake04},
more likely to be attributed to the thick disc component
also because of its proper motion.

A different approach to the same problem
is that based on a statistical analysis of full set of events.
Within this framework in previous analyses we have considered
the set of events reported by the MACHO collaboration
and shown that, on the basis of both their 
number and their characteristics (duration
and spatial distribution), they can not
all be attributed to self lensing \citep{mancini04}.
In \cite{novati06} we have considered the possible
role played by the LMC dark matter halo, and in particular
we have challenged the current view that  the halo fraction in form of MACHOs 
for the MW and the LMC are equal.
In \cite{novati09b} we have discussed
the results of the OGLE-II campaign towards the LMC,
confirming in particular the outcomes of \cite{ogle09}
in that the observed rate was consistent with the
expected self-lensing signal. 
Finally, \cite{mancini09} has reconsidered microlensing
towards the LMC by adopting a non-gaussian velocity
distribution for the sources.

Here we report on a detailed analysis of the recent
results of the OGLE-III campaign towards the LMC
\citep{ogle11}. In particular we estimate the number
of the expected events for all the possible lens population
with the purpose to derive an accurate limit
on the halo fraction in form of MACHOs.
As a result, overall, we find the observed rate to be compatible
with the expected self-lensing signal.
The plan of the paper is as follows.
In Section~\ref{sec:ogle} we resume the main
outcomes of the OGLE analysis, with in particular 
a discussion on the issue of blending
(and a comparison with the MACHO and the EROS strategies).
In Section~\ref{sec:analysis} we present our analysis.
After introducing, Section~\ref{sec:rate}, our mail tool of investigation,
the microlensing rate, and the assumed astrophysical model, Section~\ref{sec:model},
in Section~\ref{sec:nexp} we derive the expected
microlensing quantities, number of events and duration.
On the basis of this result, in Section~\ref{sec:nat}
we discuss the possible nature of the reported
observed events and in particular we evaluate
the limits on dark matter in form of compact halo objects.
Finally, in Section~\ref{sec:end} we present
our conclusions.

\section{Microlensing observations towards the LMC:
The OGLE campaigns} \label{sec:ogle}

\begin{figure}
\includegraphics[width=84mm]{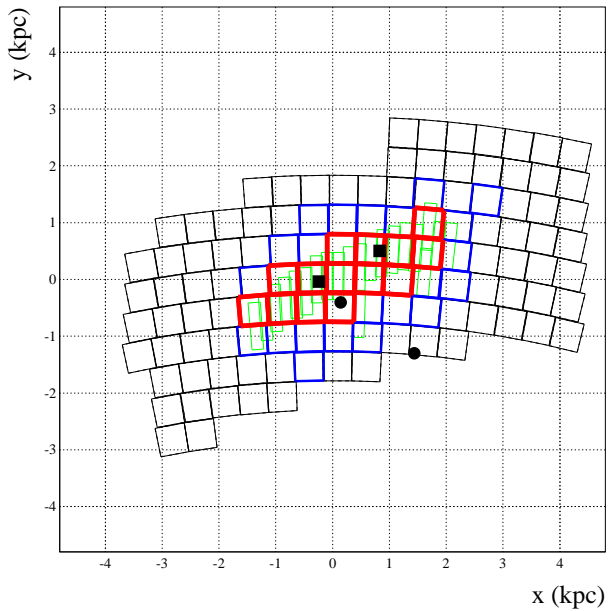}
\caption{The field of view of the OGLE-II (thin 
lines in the innermost region) and OGLE-III campaigns towards the LMC.
For OGLE-III the different contours are related to the number of monitored stars (see text for details),
moving from the more crowded inner region to the more sparse stellar fields
in the outer region of the LMC. The squares and the dots indicate
the position of the reported microlensing candidates for the OGLE-II and OGLE-III campaigns,
respectively (in particular, OGLE-LMC-03 is located at the boundary
of the monitored region). The $x-y$ reference system has its origin at the centre of the LMC,
the $x$-axis antiparallel to the right ascension
and the $y$-axis parallel to the declination.
}
\label{fig:pcampi}
\end{figure}

OGLE has continuosly upgraded its 
set up moving on successively to OGLE-II, OGLE-III
for finally recently entering in its OGLE-IV phase
of evolution. In Fig.~\ref{fig:pcampi}
we show the contours of the monitored fields of view
towards the LMC for the OGLE-II and the OGLE-III campaigns
(with overall 21 and 116 fields, respectively).
Besides the observed fields, the relevant statistics
that characterize an observational campaign
are the overall duration, $T_\mathrm{obs}$;
the number of monitored stars that act
as potential sources for microlensing events, $N^*_\mathrm{obs}$
(sometimes the product of these two term is indicated
as the overall ``Exposure'', $E$).
Finally, a selection pipeline is further specified
by the maximum allowed value
for the impact parameter, $u_\mathrm{MAX}$
and the \emph{efficiency}, usually expressed as a function of the event duration,
the \emph{Einstein time}, ${\cal{E}} = {\cal{E}}(t_\mathrm{E})$.

\begin{figure}
\includegraphics[width=84mm]{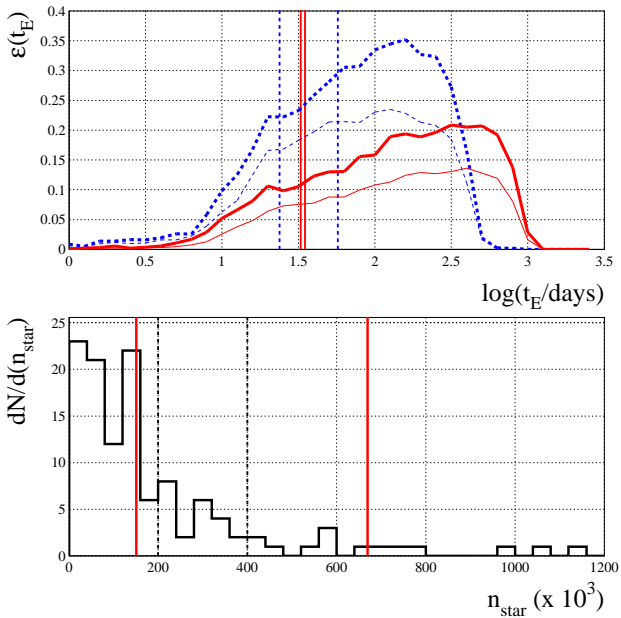}
\caption{
Top: Detection efficiency as a function of the Einstein time
for OGLE-III, field 163 (solid lines) and OGLE-II, field 4, for the Bright (bold) and
All star samples (OGLE-III field 163 partly overlaps OGLE-II field 4).  
The solid (dashed) lines indicate the reported duration
for the candidate events for the OGLE-III (OGLE-II) campaigns.
Bottom: Distribution for the number of star per field
for the 116 OGLE-III LMC fields. The solid lines indicates
the location of the fields for which the detection efficiency
has been evaluated (corresponding to the fields where the  microlensing candidate
events have been observed). The dotted lines delimit the three bins
we have used in the analysis (see text for details).
}
\label{fig:nseff}
\end{figure}

\begin{table}
\caption{Characteristics of the OGLE-II and OGLE-III
observational campaigns towards the LMC: overall duration
of the experiment, number of monitored (estimated) sources,
both Bright and All samples (see text for details), 
and overall field of view. 
}
\begin{center}
\begin{tabular}[h]{c|cccc}
&duration &  \multicolumn{2}{c}{$N^*_\mathrm{obs}$} & fov\\
&& All & Bright &\\
& (days) & \multicolumn{2}{c}{$\times 10^6$}& (deg$^2$)\\
\hline
OGLE-II & 1428 & 11.8 & 3.6 & 4.5 \\
OGLE-III & 2850 & 22.7 & 6.3 & 40\\
\hline
\end{tabular}
\end{center}
\label{tab:obs}
\end{table}
In Table~\ref{tab:obs} we report the main statistics 
for the OGLE-II and OGLE-III experimental set up
and selection pipeline (in both cases, $u_\mathrm{MAX}=1$).
We further discuss the underlying rationale for the two sets
of sources, All and Bright samples, below. In particular,
the improvement  of the available statistics given by OGLE-III
with respect to OGLE-II both in term of the experiment duration 
and number of potential sources, is apparent.
In Fig.~\ref{fig:nseff} (top) we show ${\cal{E}}(t_\mathrm{E})$ 
for the inner OGLE-III field 163, and the partly
overlapping OGLE-II field 4. As apparent, there is
a significant decrease in the OGLE-III efficiency,
in particular in the range of $t_\mathrm{E}$ values
corresponding to those of the observed candidate events (Table~\ref{tab:nobs}).
This explains, as detailed in the next Sections,
the reason of the relatively small increase in the expected
number of events for the OGLE-III campaign compared to OGLE-II.

\cite{ogle11} for their OGLE-III analysis have reported
the evaluation of the efficiency for only
the two fields of view (rather, for one CCD out of eight per each of these two fields)
where the events have been observed: field 163 and 122 located respectively in the central
and in the outer LMC region. In particular, moving towards the outer LMC regions
the fields gradually become less dense, with fewer observed objects.
A reliable measure of the crowdness of the fields 
is given by the number of estimated sources per field.
In Fig.~\ref{fig:nseff} (bottom) we show the histogram
of these values as reported by \cite{ogle11}.
\cite{ogle11} remarked
that the efficiency is larger in less crowded fields, and this turns out to be
by about a factor between 2 and 3 for field 122 with respect to field 163
(both for the All and the Bright samples of 
stars, at least for $t_\mathrm{E}>10~\mathrm{days}$, see discussion below). Accordingly,
for our analysis, we evaluate the efficiency in each field,
given the number of sources, making use of 
a linear interpolation\footnote{
The reliability of this approximation scheme is
supported by the fact that the OGLE-III
observational sampling, besides the field crowdness
a driving element of the observational setup 
for the determination of the efficiency,
was relatively uniform over all the LMC fields
(\L.~Wyrzykowski, private communication)}.
for each  value of $t_\mathrm{E}$. In the innermost and outer LMC regions, to avoid extrapolation,
we use the value of the efficiency for fields 163 and 122, 
respectively (extrapolation would in fact quickly lead to
unreasonable small values for the efficiency in the inner LMC region,
whereas in the outer region whatever choice is almost irrelevant 
as the expected rate, following the decrease in the number of sources, 
quickly drops to very small values).

For small values of the duration $t_\mathrm{E}$,
below about 5-10 days, the efficiency drops quickly below 10\%
down to 0. This makes OGLE-III, as well as OGLE-II, 
rather insensitive to MACHOs with masses below about $10^{-3}~\mathrm{M}_\odot$,
for which in particular the expected durations, not corrected for the efficiency,
are below 2~days. Furthermore,
the statistics of the simulated light curves used to evalute ${\cal{E}}(t_\mathrm{E})$
is increasingly poor moving towards smaller values of $t_\mathrm{E}$
so that all the results there should be taken with some care.  
In particular, contrary to the average result also 
reported above, the efficiency turns out to be
larger in the inner, more crowded, field 163  with respect to the outer field 122 
for $t_\mathrm{E}<5~\mathrm{days}$.

The overall large field of view monitored by OGLE-III
towards the LMC, going from the inner more crowded region
to the outer sparse field, can be exploited to
carry out analyses based on the spatial distribution.
In the following we are going to make use of this
chance, in particular we identify three bins based on the number of estimated sources per field:
the inner 14 fields with $N_\mathrm{obs}^*>400\times 10^3$ (corresponding
roughly to the region monitored by OGLE-II), the 22 ``intermediate''
fields with $200<N_\mathrm{obs}^*<400\times 10^3$ and
the outer 80 fields with $N_\mathrm{obs}^*<200\times 10^3$ (to which we refer
to also as bin 1, 2 and 3, respectively, Figs.~\ref{fig:pcampi},\ref{fig:nseff}).
 
In Table~\ref{tab:nobs} we report the informations
on the microlensing candidate events for the OGLE-II
\citep{ogle09} and OGLE-III \citep{ogle11} campaigns towards the LMC.
Comparing to our mentioned choice for the bin 
in the spatial distribution, we see that one event
is located within the inner bin (OGLE-LMC-04),
while the second event (OGLE-LMC-03)
is located within the outer bin
at the very boundary of the overall monitored fields
(Fig.~\ref{fig:pcampi}).

\begin{table}
\caption{Microlensing candidate events 
for the OGLE-II and OGLE-III
observational campaigns towards the LMC.
}
\begin{center}
\begin{tabular}[h]{c|c|c|c|c}
 event & RA & Dec & $t_\mathrm{E}$ & field \\
       & [J2000.0] & [J2000.0] & [days] & \#\\
\hline
OGLE-LMC-01 & 5:16:53.26  & -69:16:30.1 & 57.2 & 8 \\
OGLE-LMC-02 & 5:30:48.00  & -69:54:33.6 & 24.2 & 2 \\
\hline
OGLE-LMC-03 & 5:07:03.63 & -71:17:06.3 & 35.0 & 122 \\ 
OGLE-LMC-04 & 5:25:39.58 & -70:19:49.7 & 32.8 & 163 \\
\hline
\end{tabular}
\end{center}
\label{tab:nobs}
\end{table}

The choice for the two sets of source stars (and the two parallel
corresponding microlensing search pipelines), to which OGLE refers
to as the All and the Bright samples, is related to the issue of blending.
The term \emph{blending} indicates that usually more than one star is enclosed
within the seeing disc of a given resolved
object whose light curve is searched for
microlensing variations (this is opposed
to the \emph{pixel lensing} regime \citep{gould96}
where one looks for flux variations of \emph{unresolved} sources,
as for instance towards M31). Blending complicates
the analysis because, first, the real number of sources
is therefore larger than the number of observed objects;
second, it can make unreliable the estimate
of the microlensing parameters, in particular
the duration $t_\mathrm{E}$. These effects somewhat tend
to compensate one each other for the evaluation
of the optical depth, nonetheless blending remains
a main issue of concern for the interpretation
of microlensing results. Indeed, only the acknowledgement
and then a deeper understanding of this problem
lead finally to reconcile theory and observations for 
the optical depth evaluations towards the Galactic bulge,
\citep{popowski05,sumi06,hamadache06} for the final analyses 
of the MACHO, OGLE and EROS collaborations, respectively
(we also recall the previous analysis of the MOA group \citep{sumi03}
who used a Monte Carlo simulation to address the issue of blending
and we refer to \cite{novati08} for a discussion on the microlensing \emph{rate}
for Galatic bulge observations). 
The strategy adopted  to minimize blending was to take
as potential sources only a subset of the \emph{brighter} 
observed objects. In particular, for these Galactic bulge searches,
the sources were selected in a well defined region
of the color-magnitude-diagram (CMD) in order to select,
as best as possible, only \emph{Bulge} (red) clump giants.
In fact, \cite{sumi06}, as opposed to the MACHO and EROS analyses,
already stressed the extent to which blending was  relevant
also for these bright objects, an issue further analysed
in \cite{smith07}, to which we refer 
for a more through discussion.
Following the underlying rationale of these Galactic bulge analyses,
a similar strategy was then adopted also by EROS for his
final LMC analysis \citep{eros07}, with however a somewhat
looser cut within the CMD specified only with the request
for the source objects to be brighter than a given threshold, with a value
varying with the field. In the inner region the threshold magnitude
was fixed at $R_\mathrm{EROS}\sim I_\mathrm{OGLE} = 18.2$,
whereas fainter threshold values, by more than about
$1~\mathrm{mag}$, were allowed for in the outer LMC region.
This allowed to include as viable sources, besides the clump giant region, 
also, in particular, both redder 
giants and very bright main sequence stars. This way
EROS reduced the number of sources down to about 20\%
of the original set, with $N^*_\mathrm{obs}=5.5\times 10^6$ spread over its
(huge) LMC field of view of about $84~\mathrm{deg}^2$.
On the other hand the MACHO collaboration for his
final analysis of 5.7~years of data \citep{macho00}
had choosen a different strategy allowing also
for faint sources. Furthermore, MACHO
looked mostly at the inner, and denser, LMC region with
for an overall field of view of $13.4~\mathrm{deg}^2$.
The MACHO detection efficiency \citep{macho01} was then specifically designed
to hold for a total exposure of $E\equiv N^*_\mathrm{obs}\cdot T_\mathrm{obs} 
= 6.12\times 10^7$ \emph{object}-years
(rather than for stars), for $11.9\times 10^6$ observed objects,
with an estimated mean value stars (brighter than $V=24$) per object 
of $10.84\pm 2.4$.
In particular, the reported efficiency was found to vanish for
unresolved stars fainter than $V\sim 24$ (while observing objects
down to $V\sim 22$).
We recall that the difference in their respective strategies,
looking down to faint viable sources in the inner LMC region
(MACHO), rather than for bright sources only across
a much more extended field of view (EROS), has 
already been considered as a possible way out
to explain the discrepancy already mentioned above in their
results as for the compact halo objects fraction
in the mass range $(0.1-1)~\mathrm{M}_\odot$ (\cite{moniez10} and references
therein). For this reason the mentioned choice of OGLE, which
we now discuss in more detail, looks
promising to better address this issue.

The strategy adopted by OGLE for its
final LMC analyses, similar for both OGLE-II
and OGLE-III, has been to carry out
two parallel microlensing search pipelines on two distinct sets of sources.
The All sample, for objects down to a limit
threshold magnitude of $I_\mathrm{C}=20.4$;
and a Bright sample composed by a $\sim 30\%$ subset of the brighter sources
with threshold magnitude fixed at $1~\mathrm{mag}$ fainter
than the CMD red clump center ($I_\mathrm{C}=18.8$ for 
their reference field LMC\_SC1, located in the innermost LMC region). Through an analysis
of the underlying luminosity function,  OGLE estimated the number
of source \emph{stars} per observed \emph{object}
down to a magnitude $I=23.9$ (23.4) for OGLE-II (OGLE-III) respectively.
As a measure for the impact of blending, we report (Table~\ref{tab:sample}) the ratio 
of the number of stars per object 
for the three bins in the spatial distribution
already introduced. As to be expected, blending 
is more relevant in the crowded inner LMC region (bin  1), with 1.27 (1.14)
stars per object for the All (Bright) sample, respectively
(we recall that for OGLE-II the corresponding reported values
were 2.1 and 2.9\footnote{The reason behind the smaller values
reported for OGLE-III should be looked into a combination
of effects: first and more important the better
image quality of OGLE-III; then the 0.5 magnitude difference
for the threshold value, and finally the refined OGLE-III analysis
used to determine the underlying luminosity function 
(\L.~Wyrzykowski, private communication).}). In particular, we find
worth pointing out the extent to which blending
is reported to be significant also for the Bright sample, in fact
almost at the same level of the All sample
(with a relative difference of only about 10\%, both for OGLE-II
and OGLE-III). A possible reason behind this outcome
is the strategy of OGLE, its choice for the threshold magnitude values 
for the All and Bright samples,
which appears somewhat intermediate comparing
to the MACHO and EROS strategies.
In particular, the OGLE threshold magnitude
for the All sample is brighter than that of MACHO
(to compare with $V$ band values of the MACHO analysis we recall that
the color for the red clump center,
where most source are located, is $V-I\sim 1.0$),
while the OGLE threshold magnitude for the Bright sample
is fainter than that of EROS. Next observational
campaigns, in particular the ongoing OGLE-IV,
might hopefully further address this issue.

\begin{table}
\caption{Mean and rms values of the ratio of the estimated number of (estimated) real stars
over the number of observed objects, for the three bins
we have identified to trace the spatial distribution
of the OGLE-III LMC monitored region (bin 1, 2 and 3 moving
from the inner to the outer LMC region, see text for details).
We also note that the ratio of the estimated number stars,
Bright over All sample, is rouhgly costant across
the fields, though rising towards
the outer region ($0.26\pm 0.03$, $0.28\pm 0.02$ and $0.30\pm 0.03$
for bins 1, 2 and 3, respectively).
}
\begin{center}
\begin{tabular}[h]{c|cc}
Bin number& All sample & Bright sample\\
\hline
bin 1 & $1.27\pm0.08$ & $1.14\pm 0.04$\\
bin 2 & $1.13\pm0.03$ & $1.07\pm 0.02$\\
bin 3 & $1.06\pm0.02$ & $1.02\pm 0.01$\\
\hline
\end{tabular}
\end{center}
\label{tab:sample}
\end{table}

\section{Analysis} \label{sec:analysis}

\subsection{The microlensing rate} \label{sec:rate}
Microlensing observations offer two main tools of investigation
to compare observations with the expected signal
and accordingly draw conclusions on the astrophysical
issues of interest: the optical depth, $\tau$, and the microlensing rate, $\Gamma$
\citep{mao08}.
The first is the instantaneous probability to observe
a microlensing event, and therefore a \emph{static} quantity.
In particular, $\tau$ is directly related to the overall lens 
population density distribution but it turns out to be independent
from the lens \emph{mass}, one of the more crucial parameters
one is usually interested to. This is at the same time
a bonus, as it makes $\tau$ less model-dependent, and a limit,
as it does not allow one to actually characterize the observed events.
To this purpose the microlensing rate is therefore the quantity of choice
being a \emph{dynamic} quantity which allows one to compute
the number of microlensing events per unit time 
for a given number of monitored stars.
In particular, through the analysis of $\Gamma$, one can 
estimate the \emph{number} of the expected events, $N_\mathrm{exp}$,
and their characteristics, more notably their \emph{duration}
and spatial \emph{position}:
\begin{equation} \label{eq:nexp}
N_\mathrm{exp} = N^*_\mathrm{obs} T_\mathrm{obs} 
\int{\mathrm{d}t_\mathrm{E} \,\frac{\mathrm{d}\Gamma}{\mathrm{d}t_\mathrm{E}}\,
{\cal{E}}(t_\mathrm{E})}\,,
\end{equation}
where we have explicitily taken into account of the experiment detection efficiency
and written the rate in differential form (which provides also the expected 
duration distribution for the events). 

Microlensing observations, at least for the more usual
lensing configuration of single point source and point lens
with uniform relative motion, allow one the estimate of a single physically
relevant quantity, the event duration $t_\mathrm{E}$, which
is a function of the lens and source distances,
the lens mass and the modulus of the lens-source 
relative transverse velocity which relates the Einstein radius,
the microlensing event cross section, to the Einstein time,
$v_t = R_\mathrm{E}/t_\mathrm{E}$. As distances, velocities and lens mass
are not directly observable, one has to assume a model for all these quantities
to integrate them out. The details of the models we use, 
as well as details on the evaluation of the microlensing  rate,
are discussed in our previous works, and in particular we refer to \cite{novati09b}
for a discussion on all the possible source and lens population we consider:
for the sources, disc and bar LMC stars; for the lenses, disc and  bar LMC stars
(a case to which we refer in particular as ``LMC self lensing''),
Galactic disc stars and finally stars of both the LMC and the Galaxy stellar haloes,
all of these lens populations contributing to  
``self lensing''; finally, for MACHO lensing, compact halo objects
for both the MW and the LMC dark matter haloes.

\subsection{The astrophysical model} \label{sec:model}

The modelling of the LMC, in particular of its
luminous components, is an extremely live subject of research.
Our model, as described in detail in \cite{mancini04}, 
is mainly based on the work of van Der Marel and collaborators
(\cite{vdm02} and references therein), whose conclusions
have been challenged by several authors (see for instance
the recent analyses of \cite{subramanian10,bekki11} and references therein).
It is beyond the purpose of the present analyis to further
address this issue. This is because, as we detail
in the following, our main results
with respect to compact halo objects,
are almost insensitive to the exact amount of
the LMC self-lensing contribution.
The same argument applies also with respect to
the stellar haloes, with  the additional caveat
that  their modeling is even more problematic. In particular,
for the LMC stellar halo (that giving the larger contribution
of the two), here we use our \cite{novati09b} ``fiducial'' 
model based on the analysis of \cite{alves04}. As also discussed
in \cite{novati09b}, the more recent analysis of \cite{pejcha09}
suggests a different spatial distribution for this component that may indeed
significantly enhance the microlensing signal.

A fully detailed knowledge of the stellar structure of the MW is still lacking.
To  our purposes, the relevant quantitites are the local stellar density and the vertical distribution.
An  accurate modelling becomes particurarly important for a survey
as  OGLE-III where MW disc lensing can be expected, besides
LMC  self lensing, to give a relevant contribution to the rate.
As we further address in the following sections, this is because of the broader spatial distribution
of  the expected MW disc signal, beyond the inner LMC region,
with respect to that expected from LMC lenses.
For  this reason we update the model used in \cite{novati09b}.
We assume a standard double exponential disc model \citep{dehnen98}
with scale lenght $R_d$ and scale height $h_z$,
summing a thin disc and a thick disc component.
Following the detailed stellar mass budget of \cite{kroupa07}
we fix the thin disc \emph{stellar} central density to 
$\rho_{\mathrm{thin},\odot} = 0.044~\mathrm{M}_\odot \mathrm{pc}^{-3}$,
to be compared for instance with $0.038~\mathrm{M}_\odot \mathrm{pc}^{-3}$
in \cite{flynn06}.
For a scale height $h_{z,\mathrm{thin}} = 250~\mathrm{pc}$ this gives
a local disc column density
$\Sigma_\mathrm{thin} = 22~\mathrm{M}_\odot \mathrm{pc}^{-2}$
(we do not consider the luminous \emph{gas}
component that does not lens).
To complete our fiducial thin disc model we fix $R_{d,\mathrm{thin}}=2.75~\mathrm{kpc}$.
Larger values of $h_z$ are also reported. For instance,
$h_{z,\mathrm{thin}} = 300~\mathrm{pc}$ \citep{juric08}
would lead to $\Sigma_\mathrm{thin} = 26.4~\mathrm{M}_\odot \mathrm{pc}^{-2}$.
The details of the model for  the thick disc 
are less well constrained. This is important
for  microlensing as the larger scale height
of this component is expected to significantly enhance
the signal \citep{gould94,gould94c}. Here we follow the recent
analysis of \cite{dejong10} and we take as our fiducial model  
$\rho_{\mathrm{thick},\odot} = 0.0050~\mathrm{M}_\odot \mathrm{pc}^{-3}$,
$h_{z,\mathrm{thick}}=750~\mathrm{pc}$,
summing up to $\Sigma_\mathrm{thick} = 7.5~\mathrm{M}_\odot \mathrm{pc}^{-2}$,
and $R_{d,\mathrm{thick}}=4.1~\mathrm{kpc}$.
These values, also compatible with the analysis of \cite{juric08},
indicate a quite substantial thick disc over thin disc local density fraction, $f_\mathrm{thick}$,
here $f_\mathrm{thick} = 11\%$ (together with
a relatively small scale height) compared to previous analyses
where values in the range $f_\mathrm{thick} = 1\%-10\%$ were reported
(we refer to the discussion in \cite{juric08,dejong10} and references therein).
As an extreme case of a ``light'' thick disc we will report
also the results for $f_\mathrm{thick} = 1\%$.
For the thin (thick) disc component we assume
a line-of-sight velocity dispersion of $30~\mathrm{km/s}$ ($40~\mathrm{km/s}$), respectively.

\subsection{Expected events number and duration} \label{sec:nexp}
 
The starting point of our analysis is the evaluation
of the differential microlensing rate, which enters in
Eq.~\ref{eq:nexp}, towards (the center of) all the 116
OGLE-III fields. This allows us to characterize the expected microlensing signal, 
in particular the events duration and number, that we now discuss in turn.

\begin{figure}
\includegraphics[width=84mm]{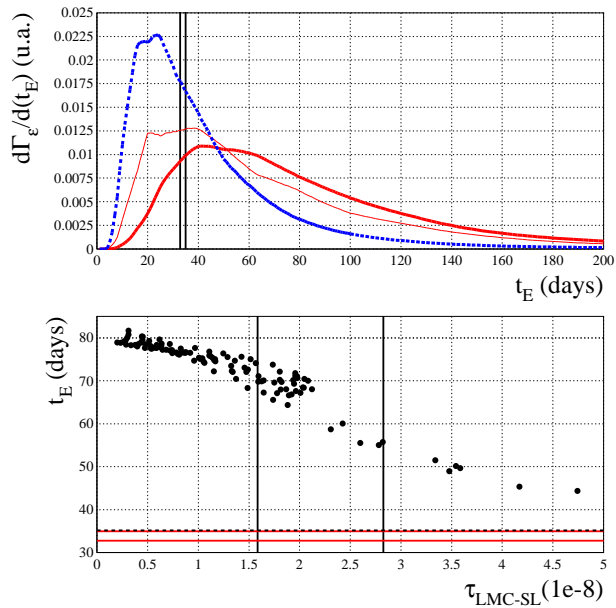}
\caption{
Top: The differential rate, $\mathrm{d}\Gamma/\mathrm{d}t_\mathrm{E}$,
corrected for the efficiency and normalized, for LMC self lensing
evaluated along the line of sight towards the OGLE-III observed events
(solid bold and thin line for OGLE-LMC-03 and OGLE-LMC-04, respectively),
and for MW disc lenses (dashed line). The vertical solid lines
indicate the values of  the duration for the two OGLE-III observed events.
Bottom: The expected median duration (corrected for the efficiency)
for LMC self-lensing events for the 116 OGLE-III fields as a function
of the corresponding value of the LMC self-lensing optical depth.
The vertical lines correspond to the optical depth values
towards the observed events, the solid horizontal lines 
indicate the values of the duration for the two OGLE-III observed events. The dotted horizontal
lines indicates the expected values for MW disc lenses. 
}
\label{fig:rate}
\end{figure}

Useful insights  can in fact be gained already by an analysis of the optical depth.
We have addressed this issue in our previous papers 
\citep{mancini04,novati09b}. Here we recall
that a main difference  for the  optical depth maps 
for MW and LMC lens populations is that
the contour levels of those for MW lenses are almost constant across the LMC field
of view. This is explained by the geometry of the configuration,
with the lenses being so close to the observer 
(a few kpc  or even less, so that the  spatial lens density distribution 
remains approximately constant  across the full monitored field of view),
and the lens distance being so small
compared to the source distance.
Coming to the microlensing rate,
this outcome translates  in particular into the
very small variations for the expected event duration 
across the LMC field of view \citep{novati09b}.
(This does not apply, however, when moving
to the expected number of events where
the source density spatial distribution,
strongly peaked at the LMC center, comes into play.)
On the other hand, the differential rate, and the expected duration, vary quite significantly
for the LMC lens populations, and this holds in particular
for the LMC self lensing, with shorter duration
expected in the inner LMC region. This is made
apparent in Fig.~\ref{fig:rate}, top panel,
where we report the, efficiency-corrected and normalized, 
differential LMC self-lensing rate calculated along the
direction towards the two observed events 
(located one in the inner and one in the outer LMC region), together with the MW disc rate. 
The plot makes also clear the  rather large variance of these distributions.
A more global picture is given
in Fig.~\ref{fig:rate}, bottom panel, where 
we report, always for LMC self lensing, the expected median duration as a function
of the optical depth (providing a stronger correlation than the distance
from the LMC center).  As a median value for the expected duration
(corrected for the OGLE efficiency) we find,
for LMC self lensing, values in the range 45-80~days,
with shorter duration expected in the central region;
for MW disc lenses (both thin and thick components), about 40~days; for compact halo
objects the duration varies with the assumed
mass. For 0.1, 0.5 and 1.0 M$_\odot$ MW MACHOs
we evaluate an expected median duration of
21, 41 and 56~days, respectively. Furthermore,
as already pointed out in \cite{novati06},
LMC dark matter halo lenses are expected to give, 
for a given MACHO mass, shorter duration
than Galactic MACHOs, with in particular $0.2~\mathrm{M}_\odot$ LMC MACHOs
expected to have similar duration than $0.5~\mathrm{M}_\odot$ MW MACHOs.

\begin{table}
\caption{Expected numer of events 
for the luminous lens populations we consider
(SL and SH stand for self lensing and stellar halo, respectively).
$N_\mathrm{exp}$ is reported as the sum over the contribution
of all the 116 OGLE-III fields (first two column) 
and the 21 OGLE-II fields (these results are reported
from our previous analysis in \citealt{novati09b}) for 
both the Bright and the All samples of sources.
}
\begin{center}
\begin{tabular}[h]{c|cc|cc}
lenses&\multicolumn{2}{c}{OGLE-III}&\multicolumn{2}{c}{OGLE-II}\\
& BRIGHT & ALL & BRIGHT & ALL\\
\hline
LMC SL        & 0.63 & 1.60 & 0.46& 1.10\\
MW disc       & 0.45 & 1.14 & 0.06& 0.12\\
LMC SH        & 0.20 & 0.51 & 0.09& 0.20\\
MW SH         & 0.09 & 0.24 & 0.03& 0.07\\
\hline                                  
total  & 1.37 & 3.49 & 0.64& 1.49\\
\hline
\end{tabular}
\end{center}
\label{tab:nexp}
\end{table}
\begin{figure}
\includegraphics[width=84mm]{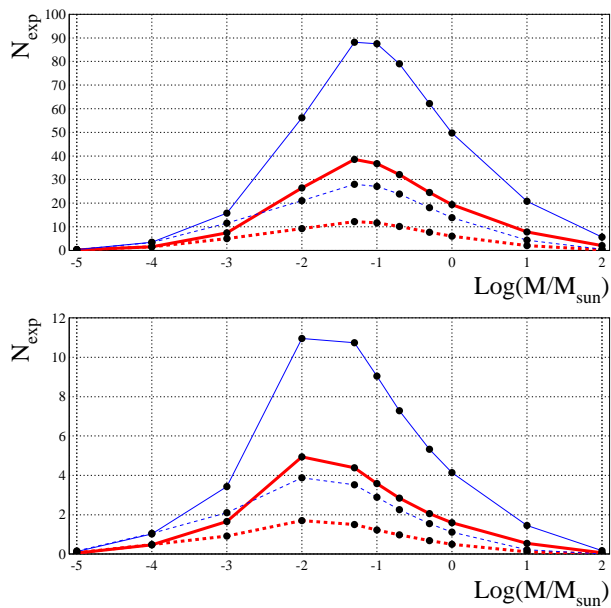}
\caption{Expected number of  MACHO lensing events
for OGLE-III (OGLE-II) campaigns (solid and dashed lines,
respectively), for both the MW (top) and the LMC dark matter haloes, 
the for the Bright (bold) and the All samples of stars. 
}
\label{fig:nexp}
\end{figure}
Moving with our analysis to the expected \emph{number} of events,
in Table~\ref{tab:nexp} we report the number of expected events
for the luminous populations considered, 
for both the All and the Bright samples of stars.
Taking into account the detection efficiency,
in Fig.~\ref{fig:nexp} we show the expected number
of events for MACHO lensing, both MW and LMC, for a set of 
delta mass function.

In  Table~\ref{tab:nexp} and Fig.~\ref{fig:nexp} we also report 
the expected number of events 
of the OGLE-II LMC campaign as evaluated in our previous
analysis \citep{novati09b}
(for MW disc lenses, using the same model
as in \cite{novati09b} we would have
found, for OGLE-III, an expected signal of 0.17 and 0.43
events, for the Bright and the All sample, respectively).
At first glance, a somewhat striking result
is a relatively overall small increase 
of the expected number of events moving
from OGLE-II to OGLE-III, when taking
into account the overall longer duration
and larger number of sources, almost a factor
of two each (Table~\ref{tab:obs}), both
entering linearly into the expression
for evaluating the expected number of events (Eq.~\ref{eq:nexp}).
As discussed in Section~\ref{sec:ogle},
the first and essential reason behind this result
should be looked into the smaller detection efficiency
reported for OGLE-III.

The expected number of MW disc lensing is 
0.45 and 1.14, for the Bright and All sample respectively
(about 70\% of the expected LMC self lensing signal), to which
the thick disc component 
contributes to about $50\%$. With reference
to the discussion on the model in Section~\ref{sec:model},
we note that this fraction would drop below $10\%$ 
for $f_\mathrm{thick}=1\%$,
(and overall the MW disc over LMC self lensing signal 
to 40\%). On the other hand,
for a thin disc scale height of $300~\mathrm{pc}$
we would get to an expected number of events,
for the Bright and All sample respectively,
of 0.54 and 1.38, almost 90\% of the LMC self lensing signal.
In any case, among the contributions to self lensing,
these values make apparent the relevance 
of the MW disc signal with respect
to the LMC self lensing signal. 

Systematic uncertainties related to the
assumed astrophysical model, besides those detailed
for that of the MW disc,
are to be expected in particular
for the other prominent luminous lensing population,
the LMC stars. We have addressed this issue
in a previous analysis \citep{mancini04}.
For somewhat extreme models of LMC, both structure and 
kinematic,  we found variations in
the expected number of events up to 30\%-50\%.

The expected number of events is related,
besides to the lens population under exam,
to the underlying source density distribution.
Together, these elements determine the
expected spatial distribution of the  microlensing events.
For a large enough set of events the spatial distribution  becomes
an extremely powerful tool of analysis
to better address an essential problem of this analysis, namely 
the understanding on the nature 
of the observed events (as we did in our previous
analyses of the MACHO events in \citealt{mancini04,novati06}).
For the present analysis of the OGLE-III results, with only 2 observed events,
we can not expect this analysis to give unambiguous outcomes.
We may however take advantage of 
the rather large overall field of view
of the OGLE-III fields across the LMC 
to at least address this issue and gain 
some insight into the problem.
As introduced in Section~\ref{sec:ogle},
based on the number of the sources, which traces the underlying
LMC populations, we address this issue by identifying 3 bins
within the monitored OGLE-III region.
Moving towards the outer LMC region, each bin
is composed by 14, 22 and 80 OGLE-III fields,
with a total fraction  of sources per bin of $42\%,\,27\%$ and $31\%$, respectively. 
The rationale behind this choice 
is that the innermost bin roughly corresponds to the
innermost LMC region (where in particular the OGLE-II fields
were distributed) whereas the two outer bins are choosen so to have, overall,
roughly the same number of sources. The strong variation
into the number of sources per field, moving from the crowded
inner region to the more sparse LMC outskirts, is clearly reflected
in the increasing number of fields per bin.
The expected spatial distribution according
to this bin choice is given
in Table~\ref{tab:nbin} where we report the fractions
of the expected number of events per bin per each lens population considered.
The outcome is driven by two competitive effects: the larger
number of sources towards the LMC centre against
the corresponding decrease for the detection efficiency.
The other aspect is related to the lens spatial distribution,
with LMC lenses more concentrated in the innermost region
and MW lenses almost uniformly distributed across the monitored
field of view.
As a result, the LMC self-lensing signal is, as expected,
more concentrated in the inner bin, where we find about 50\%
of the expected signal.
For the other lens populations, in particular for MW disc lenses, 
we find a more smooth distribution,
with roughly 30\%, 30\% and 40\% of the expected events
moving from the inner to the outer bin (the mean
number of events per field, however, is as expected 
strongly peaked in the central LMC region).

This bin-based analysis offers also a second key of
understanding on the relatively small increase
of the number of events in OGLE-III with respect
to OGLE-II. In particular we find the MW disc signal
to increase for a much larger factor (about 3)
with respect to LMC self lensing (with a relative
increase of only a factor about 1.5). The reason behind
lies into the already discussed much stronger concentration 
of LMC self lensing
in our innermost LMC bin (corresponding also roughly
to the OGLE-II fields) with respect to the more
smoothly distributed MW disc signal.

\begin{table}
\caption{Microlensing rate analysis: fraction of the expected
number of events 
for all the lens populations we consider
(SL and SH stand for self lensing and stellar halo, respectively)
for the 3 bins across the OGLE-III fields choosen
to trace the spatial distribution (bin 1, 2 and 3
moving from the inner to the outer LMC region,
see text for details).
}
\begin{center}
\begin{tabular}[h]{c|ccc}
lenses& \multicolumn{3}{c}{fraction of $N_\mathrm{exp}$}\\
&  bin 1 & bin 2 & bin 3\\  
\hline
LMC SL & 0.484 & 0.273 & 0.244\\
MW disc & 0.262 & 0.311 & 0.426\\
LMC SH & 0.340 & 0.363 & 0.297 \\
MW SH & 0.261 & 0.314 & 0.425 \\
\hline
MW halo& 0.260 & 0.315 & 0.426\\
LMC halo & 0.282 & 0.340 & 0.378\\
\end{tabular}
\end{center}
\label{tab:nbin}
\end{table}

\subsection{The nature of the observed events} \label{sec:nat}

\begin{figure}
\includegraphics[width=84mm]{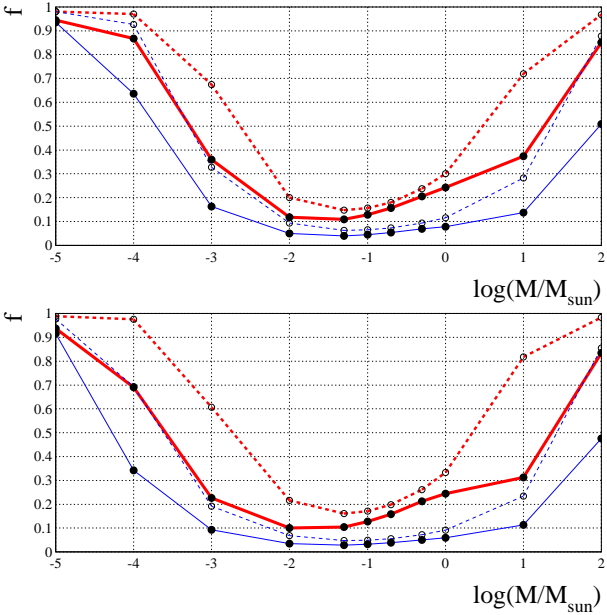}
\caption{The 95\% CL upper limit for the halo mass fraction
in form of MACHOs as a function of the MACHO mass. Bold and
thin lines for the Bright and the All samples of stars,
respectively. Dashed lines indicate the limit when
we assume the observed events \emph{are} MACHOs. Top: results of the OGLE-III campaign,
bottom: results for the joint analysis of the OGLE-II and OGLE-III
campaigns. The analysis is based upon the likelihood function
given in Eq.~\ref{eq:like}, where in particular we take
into account both the MACHO lensing and the self-lensing
populations.
}
\label{fig:like}
\end{figure}

On the basis of the analyses carried out
in the previous Section we can attemptively
draw some conclusions on the nature of the observed signal.
Both OGLE-III events, with a duration
of about 30~days, fall at the inferior limit
of the expected duration distribution for LMC self-lensing events
(Fig.~\ref{fig:rate}).
This holds in particular (because of the correlation of the expected duration with
the distance from the LMC center, Fig.~\ref{fig:rate} top panel),
for the outer event OGLE-LMC-03,
for which we find  only 13\% of the expected events
with expected shorter duration. This fraction
rises to 19\% for the inner event, OGLE-LMC-04.
The observed durations nicely match, on the other hand,
the expected values for Galactic disc events
(median value about 40~days). The observed event durations are also
in agreement with expected Galactic MACHOs, in the mass
range $(0.1-1)~\mathrm{M}_\odot$ (median value
from 21 up to 56~days). Furthermore, as already remarked, for the Galactic lens
populations, the expected duration has a rather uniform
distribution across the entire LMC field of view,
which is also in agreement with the observed values.

The analysis on the number of expected events on the other hand, Table~\ref{tab:nexp},
indicates that LMC self lensing alone
can explain the 2 reported microlensing candidates.
This holds for the All sample ($N_\mathrm{exp}=1.60$),
but also, even if this case is somewhat more contrived, for the Bright sample
(for $N_\mathrm{exp}=0.63$ we evaluate, according to a Poisson
distribution, a 13\% probability to find more than 2 events).
These conclusions are supported
by the analysis of the spatial distribution.
In fact, the strong decrease
of the LMC self-lensing rate in the outer LMC region
is at least partially compensated
by the extremely large number of observed fields.
In particular, even if almost half of the events are expected
in the central 14 LMC fields, corresponding to the inner bin
in our division (Table~\ref{tab:nbin}), still 0.40 (0.15) events are expected in the outer bin,
giving a probability of 32\% (14\%), for the All (Bright) sample, respectively, 
to observe more than 1 event into the outer bin.

The analysis on the expected number of events
from MW disc lenses, on the other hand,
supports the hint coming from the
analysis on the duration.
In fact, with $N_\mathrm{exp}=1.14$ and $0.45$
for the All  and bright sample, respectively,
MW disc lenses by themselves
could explain both reported events
almost at the same level of confidence
than LMC star lenses.
This explanation would be, however,
even more contrived for the Bright sample
(with a 8\% probability to observe
more than 2 events, a value that would
drop below 3\% assuming 
the ``light'' thick disc model).
As for the spatial distribution,
the situation is inverse with respect to the LMC self lensing case,
with now the reported candidate in the inner bin 
more difficult to be explained. Still,
we evaluate a probability of 26\% (12\%), for the All (Bright) sample, respectively, 
to observe more than 1 MW disc event into the inner bin.

All the above considerations can be made more
quantitative taking furthermore into account simultaneously
all the possible lens populations and the characteristics
of the reported candidate events through a likelihood analysis.
In particular, by Bayesian inversion, this allow us
to evaluate, for a given MACHO mass, the probability distribution for the
halo mass fraction in form of MACHO, $P\left(f\right)$
(we consider a costant prior for $f$ different from zero
in the interval (0,1)\footnote{The choice
for the upper bound $f=1$ is in fact relevant
only for the smallest value of the MACHO mass we consider,
$10^{-5}~\mathrm{M}_\odot$, and marginally $10^{-4}~\mathrm{M}_\odot$,
for which in any case the OGLE-III experiment
is not very sensitive because of the sharp drop of the detection
efficiency for small values of the duration.}).
\begin{equation} \label{eq:like}
L\left(f,m\right) = \exp\left(-N_\mathrm{exp}(f,m)\right) 
\prod_{i=1}^{N_\mathrm{obs}}\left.\frac{\mathrm{d}
{\Gamma}_{i,\mathcal{E}}}{\mathrm{d}t_\mathrm{E}} 
\right|_{t_{\mathrm{E},event}}\!\!\!\!\!\!\!\!\!\!\!\!\!\!\!\!\,,
\end{equation}
where $\Gamma_{\mathcal{E}}$ is the microlensing rate
 corrected by the detection efficiency, 
$\mathrm{d}{\Gamma}_{\mathcal{E}}/{\mathrm{d}t_\mathrm{E}}
= \mathrm{d}{\Gamma}(t_\mathrm{E})/{\mathrm{d}t_\mathrm{E}}
\cdot \mathcal{E} (t_\mathrm{E})$.
The expression reported in Eq.~\ref{eq:like} for the likelihood allows
one to take explicitily into account the observed duration
as well as the event position as the differential rate
is evaluated along the direction towards the events.
Here both $N_\mathrm{exp}$ and the differential rate 
$\mathrm{d}\Gamma_\mathcal{E}/\mathrm{d}t_\mathrm{E}$
are to be intended as given by the sum over
all the possible lens populations,
with the MACHO lensing contributions multiplied by
the factor $f$.
In principle it is possible to rewrite
Eq.~\ref{eq:like} by taking directly into
account bins in distance from the LMC center
(similarly for istance to the pixel lensing
M31 analysis discussed in \cite{novati05}).
Following the bin choice of the previous Section
we have found this analysis not to lead to 
any substantial change in the results.

The results of the likelihood analysis are reported
in Fig.~\ref{fig:like} where, for given MACHO mass, we show the limits
for the halo mass fraction on form of MACHOs, $f$.
In particular, for the Bright star sample we evaluate a
95\% CL \emph{upper} limit well below 20\% in the mass range
$(10^{-2}-2\times 10^{-1})~\mathrm{M}_\odot$, and below 10\%
up to $1~\mathrm{M}_\odot$ for the All star sample.

Self lensing, as discussed, can account
for both observed events, however, to the purpose
of the evaluation of the limits on $f$,
it must be acknowledged that it does
not have a leading role. In fact, it is
the number of expected MACHO lensing (Fig.~\ref{fig:nexp}),
at least in the MACHO mass range $(10^{-2}-1)~\mathrm{M}_\odot$,
as compared to the number of reported  microlensing candidate,
that leads to firmly exclude a significant contribution
of this population. This is apparent by inspection
of Fig.~\ref{fig:like} (dashed lines) where we show
the results of the likelihood analysis 
assuming both
the observed events \emph{are} MACHOs: the increase
with respect to the case where we include also self lensing
is of a few percent only. In that case it makes sense, in principle,
to evaluate also a \emph{lower} limit for $f$, which we find
to vary in the range $f\sim (2-4)\%$ in the MACHO mass range
$(10^{-2}-1)~\mathrm{M}_\odot$ (for the Bright star sample).

Finally, we perform a joint likelihood analysis for 
the OGLE-II and OGLE-III campaigns (Fig.~\ref{fig:like},
bottom panel). As a result we find the upper limits on $f$,
almost unchanged (with in particular 
a marginal decrease for the All sample of stars).
Comparing to the results discussed in \cite{novati09b}
this reflects the larger statistics achieved during
the OGLE-III campaign compared to the OGLE-II one.

Our results compare well, both qualitatively
and quantitatively, with those reported
in the OGLE-III \cite{ogle11} $\tau$-based analysis.
They rule out sub-solar MACHOs  and in particular
they report an upper limit 
$f<7\%$ at 95\%CL for $0.4~\mathrm{M}_\odot$
for the All sample, a result which is almost
identical with that we reach in our analysis.

An exception to this agreement is found, on the other hand,
in the mass range above $1~\mathrm{M}_\odot$,
in particular for the All star sample,
where we estimate a relatively strong \emph{upper} limit
for the halo mass fraction, (8\%, 14\% and 51\% for 1, 10
and $100~\mathrm{M}_\odot$ respectively, at 95\% CL).
The driving motivation for the difference with \cite{ogle11} can be traced back
to our larger estimate of the expected self-lensing signal,
3.49 events, to be compared with the value 2 assumed by \cite{ogle11}
on the basis of the number of reported candidates.
The larger expected self-lensing signal drives, within the framework of a confidence
level estimate for a Poisson distribution with
a background \citep{feldman98},
to smaller values for the upper limit on the mass halo fraction
(the background signal for MACHO lensing
being that of self lensing). A second reason
is the likelihood analysis we carry out where, in particular,
besides the number of the reported candidate events one takes into account also their
\emph{duration}. Given the observed values, about 30 days,
this becomes more and more important moving
towards large values for the MACHO mass, for which
the expected durations are extremely large (for
$100~\mathrm{M}_\odot$ we evaluate a median duration
of almost 500~days).

\section{Discussion} \label{sec:end}

In this paper we have analysed the results
presented in \cite{ogle11} for the OGLE-III
microlensing campaign towards the LMC.
In particular, going beyond the optical depth-based
analysis of \cite{ogle11}, but still in agreement with
their conclusions, we have presented an analysis 
focused on the estimate of the expected number and duration
of events  through the evaluation of the microlensing rate 
for all the possible lens populations.
In particular, for MACHO lensing both the MW and the LMC dark matter halo, 
and for self lensing, LMC disc and bar, MW disc and MW and LMC stellar haloes.
As a main result we find that compact halo objects might
contribute only to a negligible fraction of the dark matter haloes.
In particular we evaluate a 95\% CL \emph{upper} limit
for $f$, the halo mass fraction in form of MACHOs,
in the range 10-20\% for values of the mass $(10^{-2}-0.5)~\mathrm{M}_\odot$,
and $f=24\%$ for $1~\mathrm{M}_\odot$
(below 10\% in this full range,
and in particular below 5\% for $(10^{-2}-0.1)~\mathrm{M}_\odot$)
for the Bright (All) samples of source stars.
Indeed, the expected self lensing turns out to be sufficient
to explain the observed signal, $N_\mathrm{ev}=2$ (both for the Bright and the All samples),
with 1.37 (3.49) expected events for the Bright (All) sample, respectively. 
However, the number of the reported  microlensing candidate events
is so small compared to the number of expected MACHO lensing events,
about 40 (90) for $0.1~\mathrm{M}_\odot$ MACHOs for the Bright (All) sample
of stars, respectively, 
that the limits on the halo fraction would not change much even assuming the events \emph{are} MACHOs.
A interesting outcome of the present analysis is the relatively
significant upper limit on the halo mass fraction
we obtain in the mass range above $1~\mathrm{M}_\odot$, at least
for the All sample (8\%, 14\% and 51\% for 1, 10
and $100~\mathrm{M}_\odot$, respectively). This is a stronger constraint
with respect to those reported in previous microlensing analyses
\citep{macho01bh,eros07} and also in the OGLE-III analysis of \cite{ogle11},
and fills the gap, from the microlensing side, with the upper limit 
from the analysis of halo wide binaries \citep{yoo04b}. Stressing
the caveat that this outcome holds for the All sample only,
we also recall the event reported towards the SMC, OGLE-2005-SMC-001,
for which there is evidence in favour for the lens being a heavy mass
(about $10~\mathrm{M_\odot}$) compact halo object \citep{dong07}.
Finally, we have also evaluated the limits for the joint OGLE-II and OGLE-III analyses,
in particular for the Bright sample of stars they remain almost unchanged.

A further relevant result of the present analysis
is the role played by the MW disc population,
for which we find the expected lensing signal
to be almost as large as that of LMC self lensing,
with a 50\% contribution from the thick disc component.
In particular we evaluate $0.45$ and $1.14$
expected events, for the Bright and the All sample
respectively (against $0.63$ and $1.60$
for LMC self lensing). Comparing to the two reported
candidate events, both the LMC and the MW disc stars
might explain the observed signal. In fact,
for the All sample, the observed
rate is smaller, though still fully compatible, than the overall expected self lensing
signal ($3.49$ events including the stellar halo components).

These results, although they look quite conclusive
on the  MACHOs contribution, at least for the mass range $(0.1-1)~\mathrm{M}_\odot$, 
still leave a few open issues.  
First, the statistics of the observed events is still
too small to carry out a meaningfully analysis
on their observed characteristics, as duration
and spatial distribution, so to draw stringent
conclusions on the exact nature of the lenses.
The observed durations, around 30~days, are 
shorter than the averaged expected durations for LMC self lensing,
which could still however, based on their expected number,
explain both the observed events. At least one event, however,
both for its position, in the outer LMC region, and duration, looks more
easily explained by the Galactic disc population.
Durations and positions are on the other hand also
compatible with MACHO lensing, especially in the mass range
$(0.1-1)~\mathrm{M}_\odot$; furthermore, the spatial
distribution might be suggestive of an asymmetry
compatible with that expected by the LMC dark matter 
halo \citep{novati06}. Again, however, the overall statistics of events
is too small to further address this issue.

The number of reported candidates is the final outcome 
of the selection pipeline.
It is beyond the purpose of this paper to address
this issue, still we stress the potential difficulty
within the evaluation of the detection efficiency
to correctly take into account the risk
of excluding bona fide microlensing candidates
(as with the $\chi^2$ cut which severely reduces the number of viable candidates
in the OGLE analyses). 
To better address this issue, whose importance is
strongly enhanced by the very small number of candidate events reported,
we believe that a full achromatic-based 2-bands analysis
(even though complicated by the blending effect)
would greatly help to more easily distinguish 
microlensing from intrinsic variables, and hopefully
to lead more easily to a larger set of reliable candidate events. After the
OGLE-II and OGLE-III campaign, for which essentialy
the pipeline was carried out with $I$-band data only,
we hope the ongoing OGLE-IV may therefore adopt this strategy.

For both its OGLE-II and OGLE-III LMC campaigns OGLE choosed
to present its results for both a, smaller, ``Bright'' and
a ``All'' sample of source stars.
The underlying reason for this choice 
is linked to the central issue of \emph{blending}.
OGLE-II reported 0 and 2 events out of the Bright and All sample selection
pipelines, respectively, whereas OGLE-III reported
2 events for both  samples. The expected self-lensing rate
is of 0.64 (1.49) for OGLE-II and 1.37 (3.49)
for OGLE-III, for the Bright (All) sample, respectively.
Once again, the overall small statistics is too small, 
however it is clear that this two-sample strategy might be taken as an opportunity
for a strong internal self-consistency check
of the full analysis (selection pipeline, efficiency and
evaluation of the number of sources). 
Here again, the ongoing OGLE-IV campaign has the chance of going beyond
these previous analyses. The obvious
strategy for enhancing the microlensing rate is to
allow for fainter sources, in particular within the All sample,
even at the risk of complicating the blending analysis. 

Finally, we have discussed the strategy of the OGLE-II and OGLE-III
campaigns as compared to the previous ones carried out
by the MACHO and the EROS collaborations. Given the caveats
discussed above, it is clear that the results obtained by OGLE,
in agreement with those of EROS \citep{eros07} do not leave
much place to a significant compact halo objects population.
It remains increasingly difficult therefore to explain, within this framework,
the results of MACHO \citep{macho00,bennett05}. A possible way out
is, once more, that of a significant increase in the expected
statistics for self-lensing events, so to allow one
to use them as a strong test case to be compared with
the expected MACHO lensing signal. In this respect
the ongoing OGLE-IV campaign \citep{udalski11}, with its still increased
overall field of view, the MOA campaign
\citep{sumi11}, as well as SuperMACHO \citep{rest05},
might hopefully help to definitively address at least the issue
of the contribution of faint compact objects 
and fill the gap between the MACHO and the EROS strategies
and results.

\section*{Acknowledgments}
We are grateful to the referee for interesting suggestions and remarks
that helped us to improve the manuscript.
We warmly thank \L.~Wyrzykowski for carefully reading the manuscript,
useful discussions and comments as well as for making
available to us some unpublished data of its analysis. 
We acknowledge support by MIUR through PRIN 2008 
prot.~2008NR3EBK.

\bibliographystyle{mn2e}

\bibliography{ogle11} 

\label{lastpage}

\end{document}